\RequirePackage{lineno}
\documentclass[preprint,12pt]{elsarticle}

\usepackage{graphicx}
\usepackage{amssymb}
\usepackage{bm}        % for math
\usepackage{txfonts}

\journal{Physics Letters B}

\begin{document}
\begin{frontmatter}

\title{First determination of an astrophysical cross section with a bubble chamber: the $^{15}$N($\alpha$,$\gamma$)$^{19}$F reaction}

\author{C.~Ugalde}
\address{Department of Astronomy and Astrophysics, University of Chicago, Chicago, Illinois 60637, USA}

\author{B.~DiGiovine}  
\author{D.~Henderson}
\author{R. J.~Holt}
\author{K. E.~Rehm}
\address{Physics Division, Argonne National Laboratory, Argonne, Illinois 60439, USA}

\author{A.~Sonnenschein}
\address{Fermi National Accelerator Laboratory, Batavia, Illinois 60510, USA} 

\author{A.~Robinson}
\address{Department of Physics, University of Chicago, Chicago, Illinois 60637, USA}

\author{R.~Raut\footnote{Present address: UGC-DAE Consortium for Scientific Research, Kolkata Centre, Kolkata, India.}}
\author{G.~Rusev\footnote{Present address: Chemistry Division, Los Alamos National Laboratory, Los Alamos, New Mexico 87545, USA.}}
\author{A. P.~Tonchev\footnote{Present address: Physics Division, Lawrence Livermore National Laboratory, Livermore, California 94550, USA.}}
\address{Department of Physics, Duke University, Durham, North Carolina 27708, USA}
\address{Triangle Universities Nuclear Laboratory, Durham, North Carolina 27708, USA}

\begin{abstract}
%% Text of abstract
We have devised a technique for measuring 
some of the most important nuclear reactions in 
stars which we expect to provide considerable 
improvement over previous experiments. Adapting ideas from dark 
matter search experiments with bubble chambers, we 
have found that a superheated liquid is 
sensitive to recoils produced from $\gamma$ rays 
photodisintegrating the nuclei of the liquid. The main
advantage of the new target-detector system is a gain 
in yield of six orders of magnitude over 
conventional gas targets due to the higher
mass density of liquids. Also, the detector 
is practically insensitive to the $\gamma$-ray
beam itself, thus allowing it to detect only the 
products of the nuclear reaction of interest. 
The first set of tests of a superheated target with a 
narrow bandwidth $\gamma$-ray beam was completed 
and the results demonstrate the feasibility of the scheme.
The new data are successfully described by 
an R-matrix model using published resonance parameters.
With the increase in luminosity of the next generation 
$\gamma$-ray beam facilities, the measurement of thermonuclear 
rates in the stellar Gamow window would become possible. 
\end{abstract}


\end{frontmatter}

\section{Introduction}
Thermonuclear burning in stars is responsible for the synthesis of most 
of the chemical elements heavier than lithium in the universe. This is a 
highly temperature dependent process in which progressively heavier nuclei 
are produced as the increasingly strong Coulomb barrier is
overcome at higher and higher temperatures. 

The determination of the reaction rates of some relevant nuclear processes 
is one of the leading problems in stellar structure, evolution, and 
nucleosynthesis \cite{Woosley:2007}, with experimentally measured values 
always preferred over theoretical predictions. However, since experiments 
at astrophysical energies involve minute cross sections and thin targets, 
most current determinations of reaction rates are performed by combinations of 
experiment and theory to various degrees. Here, we describe a novel technique 
using thick liquid targets that will be useful for measuring some of the most 
important nuclear reactions at energies relevant for stellar environments. The 
sensitivity of this technique is six orders of magnitude higher than that of 
some of the most sensitive direct measurements performed to date. 

\section{Method}
The new method is based on two principles: the reciprocity theorem for 
nuclear reactions, which relates the cross sections of forward and time-inverse 
nuclear processes; and the ability of a superheated liquid to induce nucleation
when exposed to radiation \cite{Glaser:1952}. Reciprocity allows one to deduce 
the cross section $\sigma_A$ for particle capture (X,$\gamma$) processes to the 
ground state by measuring the cross section $\sigma_B$ for photodisintegration ($\gamma$,X) 
reactions, i.e.
\begin{equation}
\label{reciprocity}
{\omega_A}{{\sigma_A(X,\gamma)}\over{\lambdabar_\alpha^2}}={\omega_B}{{\sigma_B(\gamma, X)}\over{\lambdabar_\beta^2}},
\end{equation}
where X is the captured particle, $\lambdabar_\alpha$ and $\lambdabar_\beta$ 
are the channel wavelengths for capture and photodisintegration, and $\omega_A$ 
and $\omega_B$ are their respective spin factors. In the energy regimes discussed 
here, the transformation factor can provide a gain of over two orders of magnitude 
in cross section. The advantage of this principle has been used widely in the past 
in Coulomb dissociation experiments \cite{Baur:1986}. 

\begin{figure}
\includegraphics[width=5.5in]{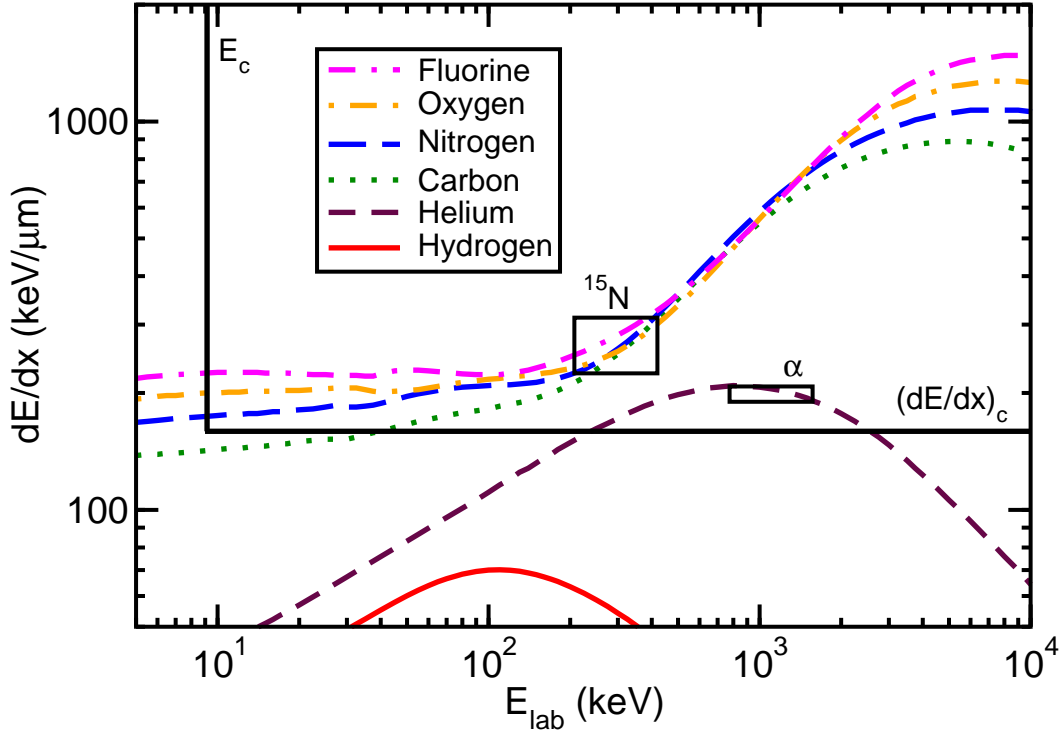}
\caption{\label{fig:stoppingC4F10}
(Color online). Nucleation thresholds for liquid C$_4$F$_{10}$ at
P = 150 kPa and T = 310 K. The curves are the stopping powers of the 
most abundant ions present in the liquid. The acceptance window is 
delimited by the black solid lines. The horizontal black line is the 
stopping power threshold (dE/dx)$_c$ and the vertical black line is 
energy threshold E$_c$. Particles above (dE/dx)$_c$ and to the right 
of E$_c$ will induce nucleation while others remain undetected. The 
small black boxes represent some of the relevant energy and stopping 
power combinations for ions from $^{19}$F($\gamma$,$\alpha$)$^{15}$N 
at $\gamma$-ray energies between 5.0 and 6.0 MeV.
}
\end{figure}

Capture reactions such as ($\alpha$,$\gamma$), (p,$\gamma$), and (n,$\gamma$), 
are responsible for many nucleosynthetic processes occuring in stellar 
environments. If the reaction product of the (X,$\gamma$) process is
a stable isotope then it can be studied experimentally by photodisintegration
if a suitable target of the reaction product can be produced. For example, the
$^{12}$C($\alpha$,$\gamma$)$^{16}$O and $^{15}$N($\alpha$,$\gamma$)$^{19}$F 
reactions can be studied via photodisintegration with 
$^{16}$O($\gamma$,$\alpha$)$^{12}$C \cite{Gai:2010} and $^{19}$F($\gamma$,$\alpha$)$^{15}$N 
using targets of oxygen or fluorine containing compounds, respectively. When nuclei are
photodisintegrated, the residual particles acquire an energy equal to
the photon energy minus the Q-value of the reaction. A limitation of the
method is that it is not sensitive to capture into excited states of the 
residual nuclei.

In the case of the experiments discussed here, the residual particles were 
detected with a bubble chamber. Originally invented for particle physics experiments, 
superheated liquid detectors have recently found new applications in several dark matter 
searches \cite{Zacek:1994,Girard:2005,Bolte:2007}. Here, we 
describe the first application of such a detector for low count rate
experiments in nuclear astrophysics.

A particle moving in a liquid deposits energy along its track until it
stops. If enough energy is deposited in a short distance, the liquid 
will be vaporized and a critical size bubble will be formed. Bubbles 
larger than a critical size will grow uncontrolled and will become visible. 
The threshold conditions for forming visible bubbles are functions of the 
degree of superheat of the liquid. Therefore, it is possible to tune the 
sensitivity of the detector to reject some minimum ionizing particles, while 
making it sensitive to heavy ions. Also, such a detector is insensitive to 
a $\gamma$-ray beam at least at a level of one part in 1$\times$10$^{9}$ \cite{Bolte:2007}. 

Fig. \ref{fig:stoppingC4F10} provides an example of stopping power curves for
ions in liquid C$_4$F$_{10}$ \cite{Ziegler:2009}. Electrons, neutrons, 
and $\gamma$ rays do not deposit energy that can trigger the bubble
chamber directly. However, these particles may transfer their momentum to other
ions by scattering interactions. In that sense, while insensitive to neutrons, 
bubble chambers can be triggered by them when they elastically scatter from
nuclei in the superheated liquid. Neutrons are very useful in the calibration of the detection
thresholds. However, they are also unwanted sources of background. The dE/dx threshold
condition for C$_4$F$_{10}$ is very sharp, with a transition slope from no 
nucleation to full nucleation of only a few keV/$\mu$m, reaching a full nucleation 
efficiency of 100\% \cite{Zacek:1994}. 

The selection of the liquid to be used in the bubble chamber depends on several 
factors. Foremost, the molecular content of target isotopes whose photodisintegration
cross section needs to be determined has to be maximized. Other isotopes present in
the molecule of the liquid may be sources of background unless reactions involving 
them have Q values above the $\gamma$-ray beam energy. Ideally, pure targets are 
desirable. However, even in very pure liquids, trace contaminants always exist, or the operating 
pressure and temperature conditions of the pure target in liquid form may be 
too extreme to work in a practical device. This is why, usually, the liquid of 
choice consists of more than one isotopic species. In principle, all liquids
should nucleate in bubble chambers \cite{Glaser:1960}. It is a matter of 
convenience to select materials that are liquid at normal pressure and 
temperature conditions. Transparent liquids are also a convenient choice 
as optical imaging techniques can be applied to detect the bubble events and 
trigger the pressure system that stops bubble growth and vaporization of the 
entire liquid volume. 

\section{Experiment}
For a proof of principle experiment we selected the $^{15}$N($\alpha$,$\gamma$)$^{19}$F 
reaction studied via the time inverse $^{19}$F($\gamma$,$\alpha$)$^{15}$N process. 
This process is the last link in the thermonuclear reaction chain leading to
the nucleosynthesis of fluorine in Asymptotic Giant Branch (AGB) and Wolf-Rayet 
stars \cite{Ugalde:2005}. The choice of reaction to study was determined 
by the existence of a strong J$^{\pi}$=1/2$^+$ resonance in 
$^{19}$F at E$_x$=5.337 MeV, which has been measured in direct ($\alpha$,$\gamma$)
experiments. C$_4$F$_{10}$ was chosen as the fluorine containing liquid as it becomes 
superheated at room temperature and pressure. 

Background reactions such as $^{19}$F($\gamma$,p)$^{18}$O or $^{12}$C($\gamma$,2$\alpha$)$\alpha$
are energetically forbidden in the energy range E$_{\gamma}$=5-6 MeV. Two-step processes like 
$^{13}$C($\gamma$,n)$^{12}$C followed by elastic scattering of the resulting neutron are
suppressed by the low abundance of $^{13}$C and by the small cross sections of the ($\gamma$,n)
reactions at energies close to the threshold (E$_{\gamma}$=4.946 MeV). 

The C$_4$F$_{10}$ liquid was contained in a cylindrical 
glass vessel with a length of 10.2 cm and an outer diameter of 3.8 cm. The length of the liquid 
target irradiated by the beam was determined to be 3.0$\pm$0.1 cm. The uncertainty was mainly 
determined by the position of the $\gamma$-ray beam with respect to the center of the 
target. This effect contributed a 3$\%$ systematic error in the determination 
of the measured cross sections. 

Pictures of the superheated liquid were taken at 10 ms 
intervals by two CCD cameras mounted at 90$^\circ$ relative to each other. The images were 
analyzed in real time by a computer and when a bubble was detected, the pressure in 
the glass vessel was increased within 40 ms of bubble formation from 160 kPa to 900 kPa. 
This led to a quenching of the growing bubble thus preventing a boiling runaway of the 
liquid. The size of the bubbles was typically 1 to 2 mm and their location 
could be determined to a precision better than 1 mm. 

\begin{figure}
\includegraphics[width=5.5in]{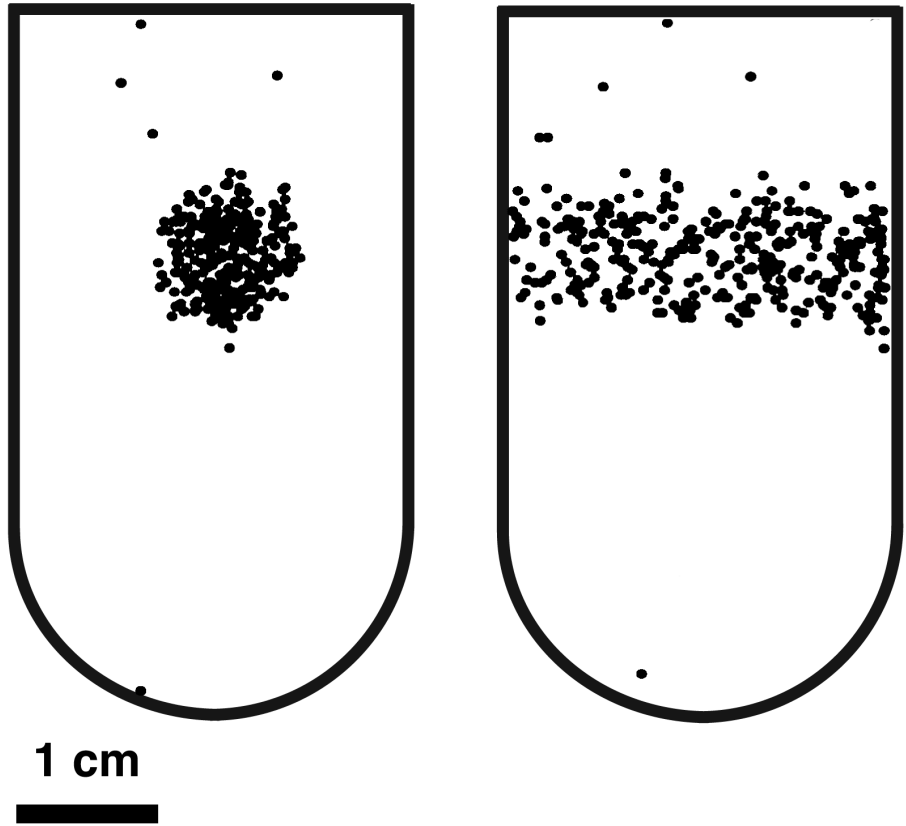}
\caption{\label{fig:fiducial}
Position of nucleation events parallel (left panel) and perpendicular (right panel) to the 
direction of the $\gamma$-ray beam as reconstructed from pictures taken by two cameras
placed at 45$^\circ$ relative to the $\gamma$-ray beam direction. The cameras were 
positioned forming a 90$^\circ$ angle between them. The beam intensity was 5.7$\times$10$^3$
$\gamma$/s and shown are events integrated over a period of 1 hour. Bubbles outside of the beam 
region correspond to background from neutrons produced either by cosmic rays or in the
walls of the experimental hall. The beam (fiducial) region contains
photodisintegration events from $^{19}$F($\gamma$,$\alpha$)$^{15}$N 
and from cosmic ray induced background. The target thickness
as seen by the $\gamma$-ray beam was determined to be 3.0$\pm$0.1 cm.
}
\end{figure}

The bubble chamber was exposed to $\gamma$ rays produced at the HI$\gamma$S facility at Duke 
University\cite{Weller:2009}. The narrow bandwidth photon beam was generated by intracavity 
Compton backscattering of free-electron-laser light from high-energy electron beam bunches. 
The photon beam was collimated with a series of three, 10 cm long, copper cylinders that 
had a 1 cm circular hole and were aligned at 0$^\circ$ with respect to the electron 
beam axis. The first collimator was located 52.8 m downstream 
from the collision point. We operated the storage ring in an
asymmetric two-bunch mode in order to reduce the beam energy 
spread. The spatial distribution of the events obtained
from the cameras correlated very well with the 1 cm diameter size
of the $\gamma$-ray beam (see Fig. \ref{fig:fiducial}).

The beam intensity was measured with a high-purity germanium detector 
positioned downstream of the target. A thick aluminum absorber 
was placed between the bubble chamber and the $\gamma$-ray detector 
in order to limit the high photon flux incident on the germanium crystal. The 
$\gamma$-ray spectrum was corrected with a Monte Carlo simulation 
of the response function of the detector and the attenuation in the 
absorber. The resulting spectrum then represents the $\gamma$-ray beam 
incident on the bubble chamber (see inset in Fig. \ref{fig:experimentTheory}).
The beam intensity ranged from 2$\times$10$^{3}$ to 3$\times$10$^{6}$ $\gamma$/s,
with a systematic error in its determination better than 5\% \cite{Carson:2010}. The beam energy  
spread was kept below 2\%. When very narrow resonances are under scrutiny, the photon beam 
resolution affects the sensitivity of the technique, as only a fraction of the beam is effective 
in exciting the resonance. This problem is also present in charged particle beam experiments. 
However, in the technique discussed here, the interaction energy spread
due to beam energy loss and straggling effects in the target is not present.

\section{Analysis and Results}

\begin{figure}
\includegraphics[width=5.5in]{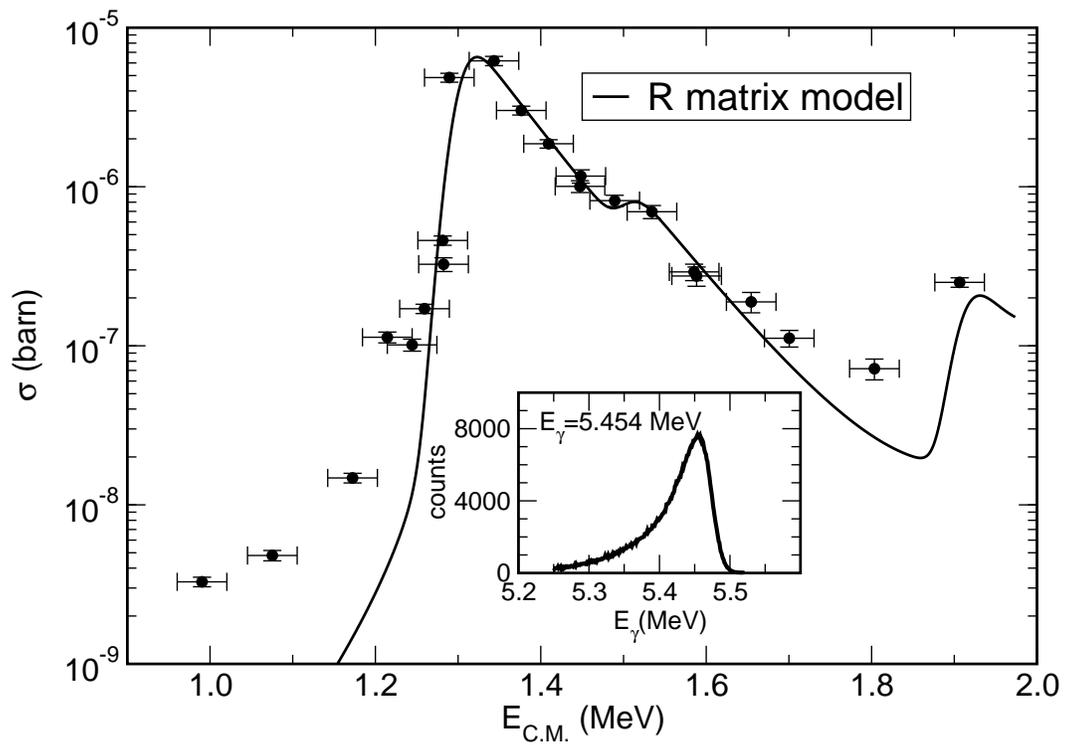}
\caption{\label{fig:experimentTheory}
Excitation function measured with a C$_4$F$_{10}$ bubble chamber 
at E$_{\gamma}$=5.0-6.0 MeV (E$_{C.M.}$=1.0-1.9 MeV). The curve 
represents a model of the $^{15}$N($\alpha$,$\gamma$)$^{19}$F 
reaction convoluted with the $\gamma$-ray beam profile. 
Solid circles represent the experimental data.
The inset
shows the energy distribution of an example of HI$\gamma$S $\gamma$-ray 
beam (centroid at E$_\gamma$ = 5.454 MeV) impinging on the bubble chamber.
}
\end{figure}

The cross section obtained from the $^{19}$F($\gamma$,$\alpha$)$^{15}$N reaction converted 
to the $^{15}$N($\alpha$,$\gamma$)$^{19}$F scale using equation \ref{reciprocity} is given 
by the solid points in Fig. \ref{fig:experimentTheory}.
The solid line in Fig. \ref{fig:experimentTheory} is the result of an R-matrix \cite{Lane:1958}
calculation performed with the AZURE code \cite{Azuma:2010}. We 
used the resonance parameters of the $^{19}$F states in the E$_x$=5-6 MeV range obtained from 
the direct $^{15}$N($\alpha$,$\gamma$)$^{19}$F measurement of \cite{Wilmes:2002} and folded 
the curve with the energy profile of the $\gamma$-ray spectrum (inset in 
Fig. \ref{fig:experimentTheory}).

For calculating the cross sections, a dead time of two seconds was determined by sampling the pressure
in the bubble chamber at a rate of 1 kHz after each event trigger. The liquid becomes
insensitive to charged particles when the superheat is removed by prompt compression.
It is then left pressurized until the bubble is quenched so that the system becomes
thermodynamically stable. The release of the high pressure superheats the liquid again. 
This pressure drop showed a gradual decline before the operating pressure and superheat 
were reached. This decline was the main source of uncertainty in the dead 
time, found to be 0.9 seconds and introducing a systematic error in the range from 
$\pm$2\% up to $\pm$15\% for measurements at the highest count rate achieved.

Two different kinds of
background sources can contribute to the bubble count rate in the detector. The first
contribution produces events that are spread evenly over the whole volume of the sensitive
liquid. The second produces events that appear in the same spatial region as the 
$\gamma$-ray beam inside the superheated liquid. The first type can be determined in a
straightforward manner by two independent methods: first, the count rate of events appearing 
outside of the beam region is compared to that of events in the path of the beam, while 
the $\gamma$-ray beam is irradiating the target. This is one 
of the reasons for which a good spatial resolution of the bubble chamber is required. In the 
experiment, this background contribution was determined to be about 8\% of the count rate 
registered outside of the beam region. This value is in good agreement with the background 
observed in a second method, where the bubble chamber was moved to the side of the beam so that the
liquid was not in the path of the $\gamma$ rays. Sources of this background are fast 
neutrons produced by cosmic rays and by the photodisintegration 
reactions in the beamline and accelerator materials that are scattered into the bubble chamber. 
The measured count rates were corrected for this background source. It would also be possible 
in the future to reduce this background contribution by passively shielding the bubble chamber 
detector with a neutron absorbing material.

The other background source cannot be easily corrected for by using the information 
from events outside of the fiducial volume. These background events are produced in 
the same spatial region as those from the photodisintegration reaction of interest. 
The main contributors to the count rate in this case are other reactions induced 
by processes such as higher-energy bremsstrahlung from collisions between the 
electrons circulating in the storage ring and the residual gas atoms in the ring. 
Assuming a vacuum of 
2$\times$10$^{-10}$ torr, an interaction length of 35 m, a beam energy of 
400 MeV, an electron beam current of 41 mA, Z=10 residual gas \cite{Schreiber:2000}, 
a 3 cm thick target with a ($\gamma$,n) cross section of 15 mb between 15-30 MeV and 0.5 mb elsewhere 
the count rate for this bremsstrahlung induced background source would be about 0.1 counts per second,
in agreement with the count rate values measured at the lowest cross sections.
This background source was studied in the experiment by varying the flux of
incident $\gamma$ rays over two orders of magnitude at an energy of 5 MeV.
The presented experimental data points have not been corrected for this
background source as further beam profiling studies need to be performed.
At an incident flux of 3$\times$10$^6$ $\gamma$/s the background allowed us 
to set an upper limit on the cross section for the
$^{15}$N($\alpha$,$\gamma$)$^{19}$F reaction at a level of 3 nb.

Neutrons produced upstream in the beam line and collimated in the
same region as the $\gamma$-ray beam are also a
possible contributor to the count rate. This set of background sources can be 
suppressed by a) choosing the threshold conditions in the bubble chamber
such that their interactions do not trigger bubble formation, b) by a 
subtraction of yields in which contaminant reactions are carefully accounted 
for, c) by placing a neutron absorber upstream in the beam line, and d) by 
identifying the neutron induced reactions through the sound they produce when 
nucleating in the superheated liquid \cite{Aubin:2008}.

The agreement between the R-matrix model and the time-inverse cross section 
measurement of the $^{19}$F($\gamma$,$\alpha$)$^{15}$N reaction is very good 
at values above 100 nb. Otherwise, the background inside the fiducial region 
dominates the measurements. Extrapolating to the highest flux that can be 
obtained at the HI$\gamma$S facility, one expects to measure cross sections 
down to 200 pb. This is a considerable improvement over the 30 nb cross section 
value that have been measured in direct experiments\cite{Wilmes:2002}, and in 
this work, where below E$_{C.M.}$=1.18 MeV, only upper limits for the 
$^{15}$N($\alpha$,$\gamma$)$^{19}$F reaction have been obtained.
A more detailed description of the experiment, including a comparison of the 
direct $^{15}$N($\alpha$,$\gamma$)$^{19}$F data with the R-matrix calculation 
will be published separately.

\section{Summary}  
We developed a new detection technique for measurements of very small cross sections which 
are of interest in nuclear astrophysics.  By studying the time reverse 
($\gamma$,$\alpha$) reaction with a $\gamma$-ray beam one can use targets which are 
thicker by six orders of magnitude. Together with an increase in cross section originating 
from reciprocity arguments, one obtains an increase in luminosity that allows a measurement 
of cross sections in the pb region even with the modest beam intensities available at 
existing facilities. Next generation facilities \cite{Luo:2010,Shimada:2010,Habs:2011} will have 
considerably higher intensities which might open the possibility to obtain experimental cross sections
for capture reactions at astrophysical energies.

We acknowledge the support of the $\gamma$-ray beam accelerator staff and scientists at Duke University. 
This work was supported by the US Department of Energy, Office of Nuclear 
Physics, under Contracts DE-AC02-06CH11357 and DE-FG02-97ER41033.

\bibliographystyle{elsarticle-num}
\bibliography{bubblePLB}

% \begin{thebibliography}{00}

% \bibitem{}

% \end{thebibliography}

\end{document}